# A literature-grounded scientific reasoning framework for defect-engineered $TiO_2$ photocatalysts

F. J. Dominguez-Gutierrez[1], E. Wierzbicka[2]
[1]National Centre for Nuclear Research, NOMATEN CoE, ul. Andrzeja Soltana 7, 05-400 Świerk, Poland
[2]Military University of Technology, Institute of Applied Physics, Solid State Physics Department, Warsaw, Poland
E-mail: javier.dominguez@ncbj.gov.pl

**Abstract.** Defect-engineered $TiO_2$ photocatalysts are extensively investigated for photocatalytic hydrogen evolution; however, the highly heterogeneous nature of the literature, including inconsistent descriptors, diverse synthesis protocols, non-uniform activity metrics, and incomplete mechanistic reporting, limits the applicability of conventional machine-learning approaches based solely on statistical regression. Here, we present a literature-grounded large language model (LLM)-assisted scientific reasoning framework for defect-engineered $TiO_2$ photocatalysts integrating curated literature data, mechanistic rule extraction, and retrieval-augmented reasoning. A harmonized database was constructed from experimentally relevant publications specifically selected for hydrogen-evolution-related defect engineering in $TiO_2$, covering polymorph-dependent behavior, hydrogenation conditions, $Ti^{3+}$ defect states, oxygen vacancies, illumination conditions, and photocatalytic activity descriptors. In parallel, mechanistic evidence sentences and publications-defined scientific rules were encoded into a structured reasoning layer enabling explainable inference beyond black-box prediction. The resulting framework combines structured experimental descriptors, semantic literature retrieval, and mechanistic interpretation to generate confidence-aware recommendations for optimal defect-engineering conditions. For example, the AI agent identified a consistent optimal anatase hydrogenation window centered at ~500 °C under $H_2$-containing atmospheres for approximately 1 h, supported by mechanistic evidence linking balanced $Ti^{3+}$/oxygen-vacancy populations with enhanced photocatalytic hydrogen evolution.

## 1. Introduction

Titanium dioxide ($TiO_2$) remains one of the most extensively investigated photocatalytic materials for hydrogen evolution owing to its chemical stability, low toxicity, earth abundance, and favorable band-edge alignment for photocatalytic redox reactions.[1,2,3] Nevertheless, pristine $TiO_2$ is limited by its wide band gap, rapid charge-carrier recombination, and low visible-light utilization, which have motivated extensive efforts toward defect engineering strategies capable of modifying its electronic and surface properties.[4,5] In particular, hydrogenation treatments,[6] oxygen-vacancy generation, $Ti^{3+}$ defect formation,[7] black/gray $TiO_2$ synthesis, and polymorph homojunction engineering have demonstrated significant potential for improving photocatalytic hydrogen evolution reaction (HER) performance.[8-13] Experimental studies have shown that moderate reduction treatments can generate beneficial $Ti^{3+}$/oxygen-vacancy populations that enhance visible-light absorption, promote charge separation, and facilitate interfacial electron transfer during proton reduction.[14-18] However, excessive defect concentrations or harsh reduction conditions may also introduce recombination centers, phase instability, or semimetallic behavior that suppresses photocatalytic activity. [18-21] Furthermore, the optimal defect-engineering conditions strongly depend on the $TiO_2$ polymorph, synthesis route i.e., reductive atmosphere, temperature, duration, gas pressure. As a consequence, the current literature on defect-engineered $TiO_2$ photocatalysts is highly heterogeneous, containing inconsistent descriptors, non-uniform hydrogen-evolution metrics, and diverse experimental protocols of PC tests that complicate direct comparison between studies and limit the applicability of conventional data-driven optimization approaches.

Recent advances in large language models (LLMs) and AI-assisted scientific agents have enabled new paradigms for knowledge extraction, literature reasoning, and hypothesis generation across materials science and chemistry. [22-25] Unlike conventional machine-learning approaches that rely primarily on numerical regression or classification, LLM-based frameworks can integrate heterogeneous textual information, contextual scientific knowledge, and mechanistic relationships extracted directly from the literature.

In the present work, we develop a literature-grounded scientific reasoning framework for defect-engineered $TiO_2$ photocatalysts focused on photocatalytic hydrogen evolution. The proposed workflow integrates a harmonized experimental database, mechanistic rule extraction, evidence-sentence retrieval, and a large language model (LLM)-assisted reasoning agent into a unified retrieval-augmented scientific platform. Relevant experimental studies on reduced and defect-engineered $TiO_2$ systems were semi-automatically curated and standardized into a structured schema containing descriptors associated with $TiO_2$ polymorphs, reduction conditions, $Ti^{3+}$ defect states, oxygen vacancies, cocatalyst configurations, illumination conditions during the PC testes, and hydrogen-evolution performance metrics. In parallel, mechanistic scientific rules and literature-grounded evidence sentences were extracted to encode relationships between defect populations, charge-carrier dynamics, and photocatalytic activity. These structured datasets and mechanistic descriptors were subsequently integrated into an LLM-assisted AI agent capable of retrieving relevant literature evidence, applying mechanistic reasoning, and generating confidence-aware recommendations for optimal defect-engineering conditions. The primary objective of this work is therefore not the development of a conventional predictive regression model, but rather the establishment of an explainable scientific reasoning framework capable of combining heterogeneous literature knowledge, structured experimental descriptors, and mechanistic interpretation for AI-assisted materials discovery and experimental planning in photocatalytic hydrogen evolution systems.

## 2. Computational Methods

The computational workflow adopted in this study is schematically illustrated in Figure 1. The workflow begins with systematic literature acquisition and semi-automatically curation of peer-reviewed publications reporting experimental HER performance of $TiO_2$-based photocatalysts under defect-engineering conditions. [26,27] Relevant experimental descriptors were extracted and standardized into a harmonized master schema containing 31 parameters associated with photocatalytic activity, including $TiO_2$ polymorphs, thermal treatment conditions, processing atmospheres, $Ti^{3+}$ defect states, oxygen vacancy characteristics, cocatalyst information, illumination conditions during photocatalytic tests, and hydrogen evolution metrics. To improve explainability and mechanistic interpretability, literature-derived evidence sentences and expert-defined mechanistic rules were incorporated into the framework. These rules encode relationships between defect engineering strategies, charge carrier dynamics, recombination mechanisms, cocatalyst effects, and photocatalytic performance. [23,26] The structured database, associated literature PDFs, and mechanistic rules were integrated into a retrieval-assisted scientific agent operating within a GPT-based retrieval-augmented reasoning environment. [26,27] The reasoning engine retrieves relevant literature evidence, applies mechanistic inference rules, integrates experimental descriptors, and generates confidence-aware recommendations for optimal defect-engineering conditions. [28] The framework outputs include suggested synthesis and treatment parameters, mechanistic interpretations, literature-supported justifications, and qualitative confidence assessments. Finally, the workflow was designed as an iterative learning architecture in which newly curated literature continuously expands the database and refines the mechanistic reasoning capability of the scientific agent.

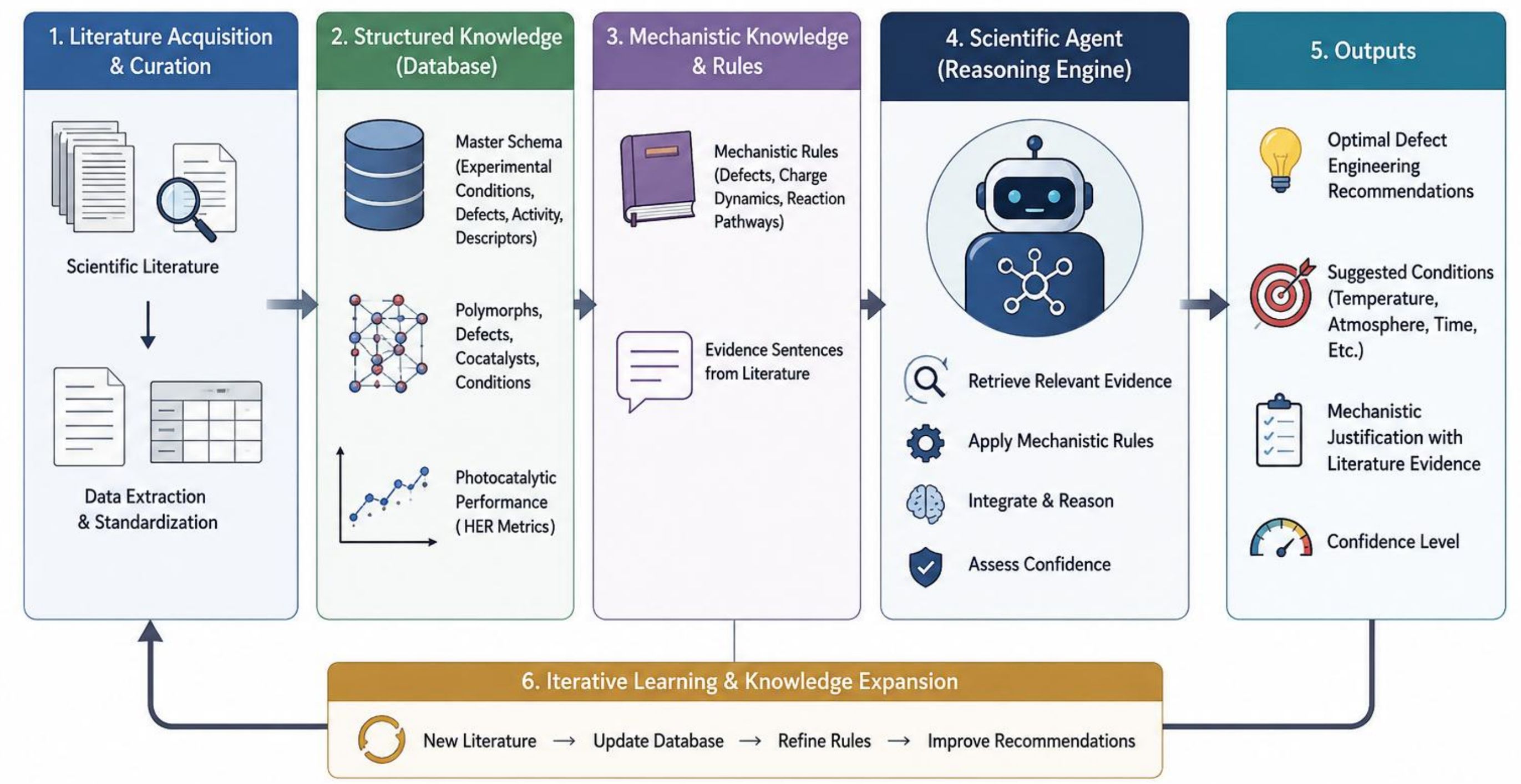

**Figure 1**. Schematic overview of the literature-grounded scientific reasoning framework developed for defect-engineered $TiO_2$ photocatalysts. The framework combines experimental descriptors, defect-related knowledge, and literature evidence to identify optimal defect-engineering conditions and continuously improves through iterative knowledge expansion.

To improve data reliability and interpretability, two descriptors were introduced: extraction tier and confidence level. The extraction tier classifies the origin and reliability of the extracted information according to the data acquisition methodology. Tier 1 (T1) corresponds to directly reported numerical data extracted from tables or explicitly stated experimental values in the manuscript text. Tier 2 (T2) refers to derived or calculated descriptors obtained through normalization, qualitative interpretation, or literature-assisted transformation of reported data. [23] Tier 3 (T3) represents values obtained through figure digitization, graphical estimation, or visually inferred information. This hierarchical classification enables transparent tracking of data provenance and extraction reliability. The confidence descriptor provides an additional qualitative assessment of data certainty within the reasoning framework. [23] High confidence entries correspond to explicitly reported experimental values or directly stated mechanistic interpretations. Medium confidence entries include visually estimated values, inferred descriptors, or partially interpreted information supported by literature context. Low confidence entries correspond to uncertain extractions, ambiguous mechanistic interpretations, or poorly resolved graphical information. The integration of extraction tiers and confidence levels allows the scientific agent to perform confidence-aware reasoning and provide explainable recommendations while explicitly communicating data reliability to the user.

The AI-assisted scientific reasoning agent was implemented using the GPT-based retrieval-augmented scientific agent through a retrieval-assisted custom agent architecture. The framework integrates the harmonized master database (`master_schema.csv`), the semantic descriptor file (`schema_description.md`), the mechanistic knowledge base (`TiO2_rules.md`), and the associated literature PDF files into a unified scientific reasoning environment. [26,27] The CSV database provides structured experimental descriptors and photocatalytic performance data, while the markdown files supply semantic interpretation and mechanistic reasoning rules required for explainable inference. Rules were retained only when supported by at least two independent literature sources or a strong mechanistic consensus. Within the custom GPT-based scientific reasoning framework, these files were uploaded into the agent knowledge base and linked through custom system-level instructions defining the operational behavior of the scientific agent. The retrieval-assisted architecture enables the agent to dynamically access relevant literature-derived information, interpret mechanistic relationships, and generate evidence-supported recommendations for optimal defect-engineering conditions in $TiO_2$ photocatalysts. The integrated PDF files allow the system to retrieve contextual scientific information directly from the original publications, improving literature traceability and mechanistic consistency. The modular architecture additionally enables continuous expansion through the incorporation of newly curated literature, allowing iterative refinement of the scientific reasoning capability of the framework.

## 3. Results

The development of the literature-grounded scientific reasoning framework required extensive standardization of heterogeneous photocatalytic literature. The final database contained 124 experimental entries extracted from 18 publications, as shown in Table 1, frequently exhibited inconsistencies in descriptor terminology, hydrogen evolution units, treatment conditions, defect classifications, and reporting methodologies. To address these limitations, a harmonized master schema consisting of descriptors was developed to systematically organize bibliographic metadata, experimental processing conditions, defect-related descriptors, photocatalytic performance metrics, illumination conditions, cocatalyst information, and mechanistic reasoning attributes. The objective of the framework is not statistical learning from large datasets, but literature-grounded scientific reasoning in domains where mechanistic knowledge is dispersed across heterogeneous publications. The selected corpus was intentionally restricted to high-quality studies reporting defect-engineered $TiO_2$ photocatalysts for HER, ensuring mechanistic consistency while enabling explainable reasoning.

Table 1. 18 peer-reviewed publications used for the scientific reasoning framework

| Paper-ID | Ref. | Paper-ID | Ref. | Paper-ID | Ref. |
|---|---|---|---|---|---|
| P001 | [4]. | P007 | [10] | P013 | [16] |
| P002 | [5] | P008 | [11] | P014 | [17] |
| P003 | [6] | P009 | [12] | P015 | [18] |
| P004 | [7] | P010 | [13] | P016 | [19] |
| P005 | [8] | P011 | [14] | P017 | [20] |
| P006 | [9] | P012 | [15] | P018 | [21] |

In table 2, we present the harmonized master schema developed in this work consisting of 31 parameters organized into seven descriptor categories to systematically represent literature-derived experimental and mechanistic information for defect-engineered $TiO_2$ photocatalysts. The first category, literature metadata (6 parameters), contains bibliographic and traceability information, including publication identifiers, journal information, and figure/table provenance used for data extraction. Experimental conditions (7 parameters) describe synthesis and treatment variables such as thermal processing conditions, treatment duration, atmospheric environment, sample type, and polymorph phase. Defect descriptors (3 parameters) encode qualitative information associated with $Ti^{3+}$ defect states, oxygen vacancy concentration, and defect localization or type. Photocatalytic performance (4 parameters) includes hydrogen evolution activity, normalized relative activity, activity classification, and associated measurement units. Illumination and reaction conditions (6 parameters) describe the photocatalytic testing environment, including light source, irradiation wavelength, light intensity, sacrificial agents, and reaction conditions. Cocatalyst descriptors (2 parameters) include cocatalyst identity and loading fraction. Finally, data quality and reasoning descriptors (3 parameters) incorporate extraction confidence, evidence sentences derived from literature, and mechanistic reasoning support, enabling explainable and confidence-aware scientific interpretation within the proposed framework.

Table 2. Summary of the 31 parameters included in the harmonized master schema used for the literature-grounded scientific reasoning framework for defect-engineered $TiO_2$ photocatalysts.

| Parameter | Description |
|---|---|
| Paper-ID | Unique identifier assigned to each literature source included in the database. |
| doi | Digital Object Identifier (DOI) of the corresponding publication. |
| Title | Title of the scientific publication. |
| year | Publication year of the study. |
| Journal | Journal where the study was published. |
| Figure or Table | Figure or table from which the experimental data were extracted. |
| Sample label | Identifier or label used for a specific photocatalyst sample within the publication. |
| Polymorph | $TiO_2$ crystal structure or phase (e.g., anatase, rutile, brookite, mixed phase). |
| Sample type | Morphology or physical form of the photocatalyst (e.g., nanopowder, nanosheets, thin film). |
| Treatment method | Experimental treatment or defect-engineering strategy applied to the material. |
| Temperature | Processing or treatment temperature in degrees Celsius. |
| time min | Duration of the treatment process in minutes. |
| atmosphere | Gas environment used during treatment or photocatalytic testing (e.g., $H_2$, Ar, Ar/$H_2$ mix). |
| H2 rate value | Numerical value of the reported hydrogen evolution activity. |
| H2 rate unit | Unit associated with the hydrogen evolution rate. |
| Relative activity | Normalized photocatalytic activity relative to the highest activity within the study. |
| activity class | Qualitative classification of photocatalytic activity (e.g., low, moderate, high, optimal). |
| Ti3 present | Qualitative indication of $Ti^{3+}$ defect concentration or presence. |
| Ti3 type | Type or localization of $Ti^{3+}$ defects (e.g., surface, embedded, mixed). |
| oxygen vacancy present | Qualitative indication of oxygen vacancy concentration or presence. |
| light source | Illumination source used during photocatalytic measurements. |
| light intensity | Light intensity during photocatalytic testing expressed in mW $cm^{-2}$. |
| wavelenght | Illumination wavelength or spectral range in nanometers. |
| sacrificial agent | Sacrificial reagent employed during hydrogen evolution experiments. |
| sacrificial agent concentration | Concentration or composition of the sacrificial agent solution. |
| cocatalyst | Cocatalyst material incorporated into the photocatalyst system. |
| cocatalyst agent concentration | Cocatalyst loading expressed in weight percentage. |

| extraction tier | Confidence tier associated with the extraction process and data reliability. |
|---|---|
| confidence | Qualitative confidence level assigned to the extracted data entry. |
| evidence sentence | Literature-derived mechanistic or experimental statement supporting the extracted data entry. |

In particular, the use of normalized relative activity descriptors allowed direct comparison between studies reporting hydrogen evolution rates in different units and under different experimental protocols. In parallel, the scientific reasoning framework developed in this work relies on complementary components: the \`schema_description.md\` file, and the \`TiO2_rules.md\` knowledge file. Where the \`schema_description.md\` file functions are built as a semantic descriptor layer defining the physical meaning, expected format, and scientific interpretation of each parameter included in the database. This file standardizes descriptor usage across heterogeneous literature sources and ensures consistency during data curation and retrieval-assisted reasoning. The schema description includes definitions for experimental parameters, defect-related descriptors, photocatalytic activity metrics, cocatalyst information, illumination conditions, and literature traceability metadata. By explicitly defining each descriptor, the framework reduces ambiguity arising from inconsistent terminology commonly encountered in photocatalytic literature. The \`TiO2_rules.md\` file contains literature-grounded mechanistic rules and experimental relationships associated with defect-engineered $TiO_2$ photocatalysts. These rules encode qualitative relationships between hydrogenation conditions, $Ti^{3+}$ defect states, oxygen vacancy concentrations, cocatalyst effects, charge carrier dynamics, recombination mechanisms, and photocatalytic hydrogen evolution activity. Each rule was additionally linked to supporting evidence sentences and associated literature references to preserve explainability and scientific traceability within the reasoning framework. The resulting rule base, as shown in Table 3, therefore acts as an expert-defined mechanistic knowledge layer that complements the structured database and enables retrieval-assisted scientific reasoning beyond purely statistical inference. In contrast to purely statistical approaches, the rule-based layer enables mechanistic reasoning and explainable recommendation generation within the scientific agent. The rules are continuously expandable as new literature is incorporated into the framework.

Table 3. Mechanistic rules extracted from the curated $TiO_2$ photocatalytic HER literature database, including rule statements, associated mechanisms, confidence levels, and descriptor keywords used within the explainable scientific reasoning framework.

| Rule ID | Title of Rule | Statement | Mechanism | Confidence | Keywords |
|---|---|---|---|---|---|
| R001 | Surface $Ti^{3+}$ defects are generally beneficial | Surface-localized $Ti^{3+}$ and oxygen vacancy defects enhance photocatalytic $H_2$ evolution activity. | Surface $Ti^{3+}/V_O$ states improve charge separation and facilitate electron transfer to adsorbed species. | High | surface defects, $Ti^{3+}$, oxygen vacancies, charge separation |
| R002 | Excessive hydrogenation causes deactivation | Long hydrogenation times or excessive temperatures reduce HER activity. | Excessive reduction promotes bulk defect formation, recombination centers, defect clustering, and phase transition. | High | overtreatment, defect clustering, bulk $Ti^{3+}$, rutile transition |

| R003 | Brookite prefers short hydrogenation times | Brookite $TiO_2$ exhibits optimal HER activity after short hydrogenation durations. | Short treatments generate beneficial surface defects without inducing structural degradation. | High | brookite, short hydrogenation, optimal window |
|---|---|---|---|---|---|
| R004 | Bulk $Ti^{3+}$ defects can become recombination centers | High concentrations of bulk $Ti^{3+}$ defects reduce photocatalytic activity. | Bulk defect states act as electron-hole recombination centers. | High | bulk defects, recombination, deep traps |
| R005 | Crystal damage can activate photocatalysis | Crystal damage and defect-rich regions can activate HER even when hydrogenation alone fails. | Crystal damage creates catalytically active surface states and charge trapping sites. | High | crystal damage, scratching, implantation, surface activation |
| R006 | High-pressure hydrogenation is highly effective | High-pressure hydrogenation can generate highly active defect-engineered $TiO_2$. | Enhanced hydrogen incorporation promotes surface $Ti^{3+}$ and oxygen vacancy formation. | Medium | HPH, hydrogenation, black $TiO_2$ |
| R007 | Surface damage alone is insufficient without suitable defects | Surface disorder without beneficial electronic defects does not necessarily improve HER. | Structural disorder may increase recombination without enhancing charge transfer. | Medium | surface disorder, recombination, defect engineering |

Figure 2 presents a representative example of the explainable AI-assisted scientific reasoning workflow implemented within the proposed $TiO_2$ scientific agent. In this example, the user queries the system regarding the optimal hydrogenation conditions for anatase $TiO_2$ photocatalysts for photocatalytic hydrogen evolution reactions (HER). The retrieval-assisted reasoning engine responds by integrating structured database descriptors, literature-derived mechanistic rules, and supporting evidence extracted from multiple curated publications. The agent identifies a hydrogenation window centered around 500 °C for approximately 60 min in an $H_2$-containing atmosphere as the most defensible general recommendation for anatase-based HER systems. Importantly, the framework does not provide only a numerical recommendation, but also generates mechanistic explanations associated with $Ti^{3+}$ defect formation, oxygen vacancy concentrations, and charge carrier dynamics. The interface further retrieves supporting evidence directly from the curated literature database, including specific papers, experimental conditions, qualitative activity outcomes, and evidence snippets used to justify the recommendation. A key feature illustrated in Figure 2 is the confidence-aware reasoning capability of the framework. Recommendations supported by multiple independent studies and consistent mechanistic trends are classified with high confidence, whereas literature-supported but less generalized conditions are assigned medium confidence. This explainable reasoning architecture demonstrates how retrieval-assisted scientific agents can transform heterogeneous photocatalytic literature into interpretable and evidence-supported experimental guidance, representing a robust alternative to purely black-box machine learning approaches for complex materials science datasets.

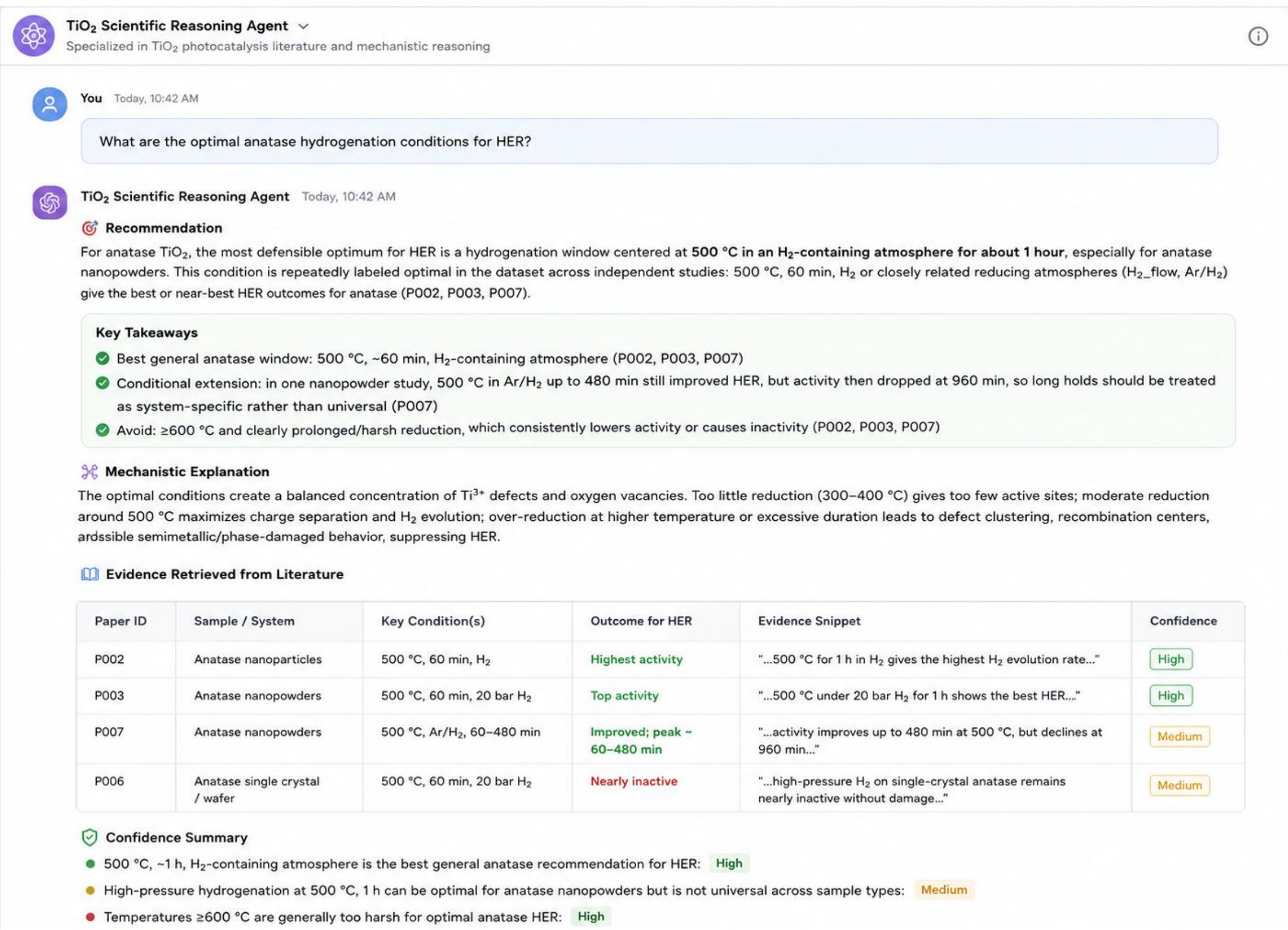

**Figure 2**. Representative example of framework operationExample of the explainable AI-assisted scientific reasoning workflow. The retrieval-assisted framework integrates structured literature data, mechanistic rules, and evidence retrieval to generate confidence-aware recommendations and mechanistic interpretations for optimal anatase hydrogenation conditions for photocatalytic hydrogen evolution reactions.

To explore descriptor relationships and identify mechanistic trends within the curated database, a temperature–time–activity heatmap was generated using hydrogenation-relevant experiments only, as shown in Figure 3. The heatmap was constructed using normalized relative activity values to enable comparison across publications employing different hydrogen evolution metrics. The clearest activity hotspot appears around 500 °C, indicating that moderate hydrogenation temperatures consistently correspond to enhanced photocatalytic performance across multiple studies. This behavior is consistent with the mechanistic trends encoded in the `TiO2_rules.md` knowledge base, where moderate defect densities and balanced $Ti^{3+}$/oxygen vacancy concentrations are associated with optimal charge separation and photocatalytic activity. Short hydrogenation treatments at 500 °C and 5–10 min exhibited strong activity particularly for brookite-based systems (P001), whereas anatase systems frequently showed optimal activity at 500 °C with longer treatments around 60 min, as observed in P002, P003, P004, and P007. Although highly active regions were also identified for longer treatment durations between 360 and 480 min, these conditions are represented by a limited number of experiments and should therefore be interpreted cautiously. Overall, the heatmap analysis provides strong evidence that approximately 500 °C represents the principal activity window for hydrogenation-based activation of $TiO_2$ photocatalysts within the current literature-grounded dataset.

**TiO2 Hydrogenation Heatmap**

Temperature vs time vs mean relative H2 activity

Hydrogenation-relevant experiments only

Untreated and non-hydrogenation controls excluded

| Treatment temperature (°C) \ Treatment time (min) | 5 | 10 | 15 | 30 | 60 | 360 | 480 | 960 |
|---|---|---|---|---|---|---|---|---|
| 300 | NA | NA | NA | NA | 0.13 (n=6) | NA | NA | NA |
| 400 | NA | NA | NA | NA | 0.05 (n=3) | NA | NA | NA |
| 500 | 0.71 (n=1) | 0.37 (n=3) | 0.42 (n=3) | 0.65 (n=2) | 0.60 (n=18) | 0.83 (n=1) | 1.00 (n=2) | 0.33 (n=3) |
| 600 | NA | NA | NA | NA | 0.12 (n=3) | NA | NA | NA |
| 700 | NA | NA | NA | NA | 0.20 (n=6) | NA | NA | NA |
| 900 | NA | NA | NA | NA | 0.02 (n=4) | NA | NA | NA |

Mean relative H2 activity: 0.0 – 0.5 – 1.0

**Figure 3.** Temperature–time heatmap of normalized relative photocatalytic $H_2$ activity for hydrogenation-treated $TiO_2$ systems extracted from the curated literature database. Cell values represent the mean relative activity and the number of experiments (n) associated with each treatment condition. The analysis reveals a dominant activity window centered around 500 °C, consistent with literature-derived mechanistic rules for optimal defect-engineering conditions.

Figure 4 presents the relationship between $Ti^{3+}$ defect concentration and normalized photocatalytic HER activity extracted from the curated literature database. The distribution reveals a clear non-linear relationship between defect concentration and photocatalytic performance, supporting the mechanistic rules incorporated into the scientific reasoning framework. Samples classified with low $Ti^{3+}$ concentrations generally exhibit limited HER activity, indicating insufficient defect formation for efficient charge separation and electron transfer. In contrast, moderate and especially high $Ti^{3+}$ concentrations are associated with substantially enhanced relative activity, suggesting that controlled defect engineering promotes beneficial surface states and improved photocatalytic performance. The highest median activity and the broadest distribution of highly active samples are observed for the high $Ti^{3+}$ category, which is consistent with the literature-derived hypothesis that optimized $Ti^{3+}$/oxygen vacancy populations facilitate charge separation and hydrogen evolution reactions. Surface-accessible $Ti^{3+}$/oxygen-vacancy states can trap photogenerated electrons productively, prolonging carrier lifetime and mediating electron transfer to water or adsorbed proton species. However, when reduction is too harsh or too prolonged, a larger fraction of defects becomes embedded in the lattice, clustered, or associated with TiOx-like reduced domains. These buried or electronically coupled defects are no longer in direct contact with the electrolyte and instead act as deep traps or recombination centers, lowering HER activity despite the larger apparent $Ti^{3+}$ population. These observations are in strong agreement with mechanistic Rules R001 and R004, which respectively describe the beneficial role of moderate surface $Ti^{3+}$ defects and the detrimental effects of excessive bulk-like defect concentrations. The boxplot representation additionally highlights the heterogeneous nature of the literature-derived dataset while preserving transparent visualization of individual experimental observations. Overall, the descriptor relationship analysis demonstrates that balanced defect engineering, rather than maximum reduction intensity, is the key factor governing optimal photocatalytic HER performance in defect-engineered $TiO_2$ systems.

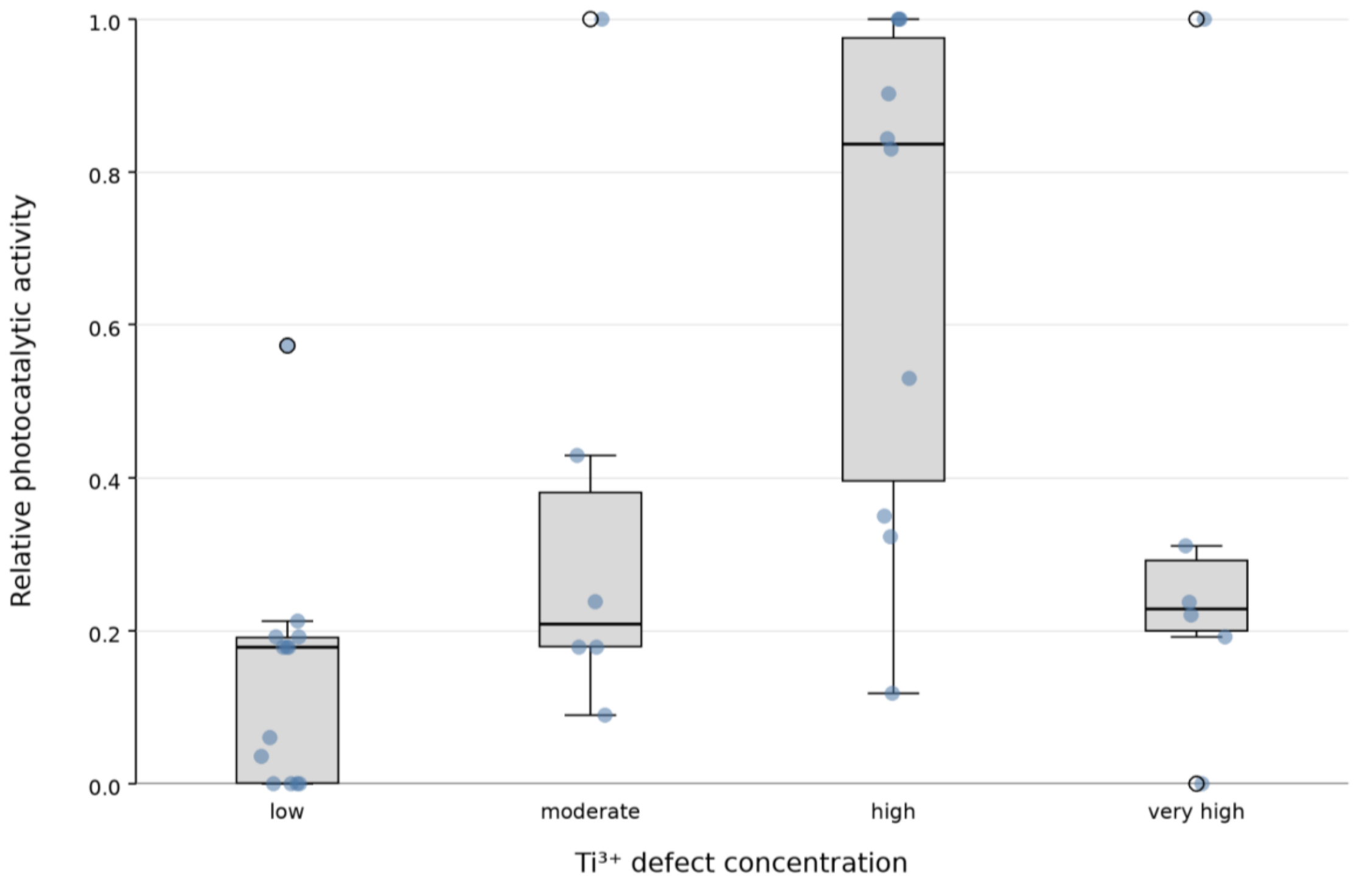

Figure 4. Relationship between $Ti^{3+}$ defect concentration and normalized photocatalytic HER activity extracted from the curated $TiO_2$ literature database.

The framework independently recovered known mechanistic trends without being explicitly programmed with optimal temperatures or activity rankings, demonstrating that literature-grounded reasoning can reproduce expert-level interpretation from heterogeneous experimental evidence. Thus, the validation results presented in Table 4 demonstrate that the recommendations generated by the literature-grounded scientific reasoning framework are fully consistent with the consensus emerging from the curated $TiO_2$ photocatalytic HER literature. For anatase $TiO_2$, the agent identified hydrogenation at approximately 500 °C for 60 min as the most robust recommendation, in agreement with independent studies reporting optimal or near-optimal hydrogen evolution performance under similar conditions (P002, P003, and P007). The framework also correctly recognized the distinct behavior of brookite $TiO_2$, recommending short hydrogenation treatments as the preferred activation strategy, consistent with the experimental observations reported in P001. Furthermore, the agent identified excessive reduction as a detrimental factor for photocatalytic performance, reflecting the consensus that prolonged hydrogenation or overly aggressive reduction conditions promote defect clustering, bulk-like $Ti^{3+}$ states, and recombination centers that suppress HER activity (P002 and P007). Importantly, these recommendations were not retrieved from a single publication but emerged from the integration of structured experimental descriptors, mechanistic rules, and evidence extracted from multiple literature sources. This agreement between AI-generated recommendations and established literature trends provides an initial validation of the framework and demonstrates its capability to perform explainable scientific reasoning while maintaining consistency with experimentally observed photocatalytic behavior.

Table 4. Validation of the literature-grounded scientific reasoning framework through representative scientific queries.

| Query | Agent Recommendation | Literature Consensus |
|---|---|---|
| Optimal anatase hydrogenation | 500 °C, 60 min | P002, P003, P007 |
| Brookite hydrogenation | Short treatment | P001 |
| Excessive reduction | Deactivation | P002, P007 |

Although the present framework demonstrates the ability to integrate heterogeneous literature data, mechanistic knowledge, and retrieval-assisted reasoning into an explainable scientific agent, several limitations should be acknowledged. First, the current database is restricted to a curated set of literature reports and therefore may not capture the full diversity of defect-engineered $TiO_2$ photocatalysts reported in the broader literature. Second, the framework relies on manual data extraction and expert-defined mechanistic rules, which may introduce subjective interpretation despite the use of confidence levels and evidence-based validation. Third, the generated recommendations are constrained by the quality and completeness of the available literature and should therefore be interpreted as literature-grounded guidance rather than definitive predictive outcomes. An additional limitation of the proposed framework is associated with extrapolation beyond well-represented regions of the literature knowledge base. To evaluate this behavior, the AI agent was queried regarding the optimal hydrogenation conditions for rutile $TiO_2$ photocatalysts for photocatalytic hydrogen evolution, despite the limited availability of direct rutile HER studies within the curated database. Rather than generating a single optimal condition, the framework identified the scarcity of direct experimental evidence and proposed a bounded screening window of approximately 450–500 °C for 10–60 min under $H_2$-containing atmospheres. The recommendation was primarily derived from mechanistic reasoning and indirect evidence obtained from hydrogenated anatase, brookite, and mixed-phase $TiO_2$ systems. Specifically, the agent recognized that direct rutile studies (P001 and P015) showed limited HER enhancement following hydrogenation, with rutile samples treated at approximately 500 °C remaining inactive or exhibiting only marginal activity improvements. Consequently, the framework avoided recommending aggressive black-$TiO_2$-type reduction strategies for rutile. The recommendation is instead supported by broader literature trends reported for anatase and mixed-phase materials (P002, P004, P007, and P014), where moderate hydrogenation treatments generated surface-accessible $Ti^{3+}$ states and oxygen vacancies associated with improved charge separation and enhanced HER activity. Mechanistic rules extracted from the literature further indicated that excessive reduction, bulk $Ti^{3+}$ accumulation, and defect clustering can promote recombination pathways and suppress photocatalytic performance. The agent therefore inferred that controlled surface reduction rather than maximum defect generation would likely represent the most promising strategy for rutile-based systems. Importantly, the framework assigned only medium confidence to this recommendation and explicitly acknowledged the absence of a literature-supported optimum for rutile HER. This example demonstrates both the capability of the framework to perform mechanistic extrapolation from heterogeneous evidence and its dependence on the coverage of the underlying literature corpus. Consequently, predictions generated outside densely populated regions of the knowledge base should be interpreted as scientifically informed hypotheses requiring experimental validation rather than definitive optimization outcomes.

## 4. Conclusions

In this work, a literature-grounded scientific reasoning framework for defect-engineered $TiO_2$ photocatalysts was developed using a retrieval-assisted large language model (LLM) agent architecture. The framework integrates a harmonized experimental database, mechanistic knowledge extraction, evidence-based reasoning, and confidence-aware recommendations to address the heterogeneous nature of the photocatalytic hydrogen evolution literature. A curated corpus comprising 18 peer-reviewed publications was standardized into a structured database containing 31 descriptors associated with $TiO_2$ polymorphs, defect-engineering conditions, cocatalyst configurations, illumination parameters, and photocatalytic performance metrics. In addition, literature-derived mechanistic rules and evidence sentences were incorporated to enable explainable scientific reasoning beyond conventional statistical data analysis. Descriptor relationship analysis identified a consistent hydrogenation activity window centered around approximately 500 °C and revealed that moderate-to-high $Ti^{3+}$ defect concentrations are generally associated with enhanced photocatalytic hydrogen evolution, whereas excessive reduction promotes defect clustering and recombination-related deactivation. The retrieval-assisted scientific agent successfully reproduced established literature trends, generating evidence-supported recommendations for optimal anatase and brookite hydrogenation conditions that were fully consistent with the consensus emerging from the curated literature database. The results demonstrate that literature-grounded scientific reasoning can effectively integrate heterogeneous experimental observations, mechanistic knowledge, and literature evidence into an interpretable decision-support framework for photocatalyst optimization. More broadly, the proposed methodology provides a scalable strategy for combining structured literature curation, mechanistic understanding, and AI-assisted reasoning toward explainable materials discovery and experimental planning in complex photocatalytic systems. The value of the framework does not lie in reproducing known trends, but in its ability to simultaneously integrate structured descriptors, mechanistic rules, and evidence from multiple publications to generate traceable recommendations in a reproducible manner.

## Acknowledgement

This work was supported by the funds granted to the Faculty of Advanced Technologies and Chemistry, Military University of Technology, Poland, within the subsidy for maintaining research potential in 2026, Grant No. UGB531000095-W900-22/2026.